\documentclass{aadebug}

\begin{document}

\corr{0309027}{269}

\runningheads{Oliver Oppitz}{A Particular Bug Trap:  ER for Virtual Machines}


\title{A Particular Bug Trap: Execution Replay Using Virtual Machines
}

\author{
Oliver Oppitz\addressnum{1}
}

\address{1}{
Winterstrasse 15, 
81543 Munich, 
Germany
}

\extra{1}{E-mail: o.oppitz@ieee.org}

\pdfinfo{
/Title (A Particular Bug Trap: Execution Replay for Virtual Machine)
/Author (Oliver Oppitz)
}

\begin{abstract}
Execution-replay (ER) is well known in the literature but has been
restricted to special system architectures for many years. Improved
hardware resources and the maturity of virtual machine technology
promise to make ER useful for a broader range of development projects.

This paper describes an approach to create a practical, generic ER
infrastructure for desktop PC systems using virtual machine
technology. In the created VM environment arbitrary application
programs will run and be replayed unmodified, neither instrumentation
nor recompilation are required.

\end{abstract}


\section{Introduction}

The concept of execution replay (ER) has been known in literature for
many years \cite{huselius_2002}. Its first step is to bring the system
into a well-defined and reproducible initial state. During the
following system execution all non-deterministic events (stimuli to
the system) like interrupts, user input, moments of scheduling
etc. are recorded. This allows to re-run the program identically in
particular for debugging purposes.

As the replay re-executes each single instruction, there is the
opportunity to analyse its behaviour in all details and single step
through it. This is in stark contrast to conventional logging
mechanisms (printf) that only record a small subset of the program's
execution. One of the biggest advantages of ER is that all
timing-dependencies are recorded during the execution phase. Therefore
it is possible to debug time-critical, multi-threaded code in
non-real-time.

Despite the simple principle, implementations of ER pose substantial
problems and until now were only available for specialized areas
e.g. for message passing system \cite{ronsse00execution}. By creating
a deterministic, replay-able virtual machine environment, it will be
possible to use execution replay for a broad range of development
projects. By using a VM emulating a complete personal computer, the
development of all software for this system can benefit from execution
replay.

The rest of this paper is organized as follows: options to implement
ER in generic (personal) computer systems are discussed in Section
\ref{sec:boundary}. Concluding that virtual machines offer many
advantages, a particular example, the User-Mode-Linux (UML) virtual
machine is introduced in Section \ref{sec:recording-vm}. Section
\ref{sec:deterministic-vm} explains how the UML VM can be adapted to
ER and gives some consideration to relevant hardware support in
current microprocessor architectures. The paper is concluded by an
outlook on promising directions for further research.

\section{Environments for Execution Replay}
\label{sec:boundary}

For implementing execution replay, a {\em system} or {\em subsystem} to
be recorded must be chosen. This can be a complete computer, an
operating system process or a module of a program. In all cases the
boundary of the system has to be defined precisely because all data
going into the system needs to be recorded in the execution
phase. During the replay this data is used to stimulate the system
identically.

For practical reasons it is difficult to record all input that a
personal computer receives: it would be necessary to attach hardware
probes which leads to problems with timing inaccuracies, signal noise
etc. Also parts of this system are not deterministic: for example the
exact timing of harddisk accesses cannot be reproduced in a replay.

Alternatively one may record the inputs to a subsystem like the CPU and
the memory system. In 1991 this was shown to work using specialized
custom hardware in \cite{bacon91hardwareassisted}. However, the
approach does not seem applicable for modern PC systems, because the
relevant signals can hardly be accessed at the main boards and --- if
possible at all --- the hardware will be expensive.

In contrast the software domain avoids many of these problems.  A
natural system to record is an OS process, because this is a unit of
processing, that is (mostly) independent from other processes.  All
data going into the process like command line arguments and file input
need to be recorded. But also interactions with the OS like signals
and system calls need to be kept track of. Unfortunately this
interface is quite complicated for real-world operating systems. Also
for parallel programs, interactions with other user or system
processes need to be recorded (including shared memory
accesses). Major adaptations to the operating system would be
necessary for this. Due to this it would be hard to guarantee a
faithful replay.

Another implementation option is to design an operating system with
execution replay in mind like the Asterix kernel
\cite{thane01asterix}. This operating system records all input to the
machine (e.g. at the driver level) and records scheduling
events. However, it is tailored for embedded systems and not
compatible with standard operating systems.

The solution proposed in this paper is a hybrid between the conceptual
simplicity of recording a complete machine and the beauty of the
software world: a deterministic, replay-able virtual machine executing
a standard operating system.

\section{Introducing the User-Mode-Linux VM}
\label{sec:recording-vm}

A virtual machine has favourable properties for execution replay. It
has a relatively simple structure, at least compared to an operating
system. As it is pure software it does not suffer from problems of the
physical domain. And last but not least it can be adapted to operate
deterministically. 

As a basis for further research and to illustrate the principle, the
open-source User-Mode-Linux VM \cite{www-dike} by Jeff Dike was
chosen. Due to its special virtualization approach, UML may only
execute Linux programs. However, there are other virtual machines like
\cite{www-vmware} emulating a complete personal computer at the
register-transfer level. These VMs can boot native operating systems
like Linux or Windows from the original installation CDs and prove
that this approach to ER is generally applicable.

The virtualization scheme used by UML is to port the Linux operating
system to a virtual UML architecture. Inside this virtual machine
there is a Linux {\it guest} kernel that executes regular, unmodified
Linux binaries.  When such a binary executes a privileged instruction
like I/O or a system call, this access is detected by the host kernel
and redirected to special handler routines in the UML binary. Also
interrupts and exceptions generated in the UML are trapped and
executed in user mode. Therefore all UML code executes in user mode
(hence the name) and no modifications to the host kernel are
necessary. The resources of the {\it host} operating system are only
accessed through (virtual) hardware drivers.  For example virtual
harddisks are mapped to files of the host system, the keyboard is
mapped to the host keyboard etc. Also network devices are supported so
that a UML can access local and remote networks.

\section{Creating a Deterministic VM}
\label{sec:deterministic-vm}

Currently UML does not support execution replay.  Therefore it is the
goal of the author to enhance it to deterministically execute, replay
and debug arbitrary software. The key to this is to record all stimuli
and their respective timing.

Input stimuli to the VM are either delivered via interrupts or via
virtual device drivers. Interrupts for the VM are implemented through
signals created by the host OS.  Like interrupts, signals are
asynchronous to the program flow of the guest processes. Therefore
during the execution phase the moments when they occur need to be
recorded. The simplest and most accurate measure for this is the
number of instructions executed until the moment when an interrupt
occurs.

In principle executed instructions can be counted in software by
instrumenting branch instructions and calculating the number of
executed instructions from the current program counter (PC). However,
this requires modifications to all executable code (kernel and
applications) which is an obstacle for regular usage.  Hardware
instruction counters solve this problem by counting the executed
(retired) instructions transparently in the background. \footnote{ In
\cite{thane00} Thane and Hansson introduce another software approach
measuring elapsed physical time with a defined resolution and
inserting breakpoint instructions into the object code. This is much
simpler than instrumenting branches, but there are (rare) cases where
the approach fails to generate a correct replay.}

A number of CPU implementations including Intel Pentium III, Pentium
IV, AMD Athlon, Itanium, and PowerPC comprise configurable counters
for counting {\it retired} instructions. Surprisingly these counters
are far from accurate. For the x86 architectures the processor
documentation states a number of cases where they count
incorrectly. In theory this can be compensated for, but tests by the
author established that there more inaccuracies that render the
counters unusable. For the Itanium no tests could be performed but the
processor documentation does not give any guarantees about correct
counter values either.

Besides the x86 implementations the Motorola PowerPC MPC7441 was
evaluated. Tests with an Apple eMac confirmed its counters to be
accurate. A minor exception are interrupts which make the counter
overcount: at each switch to and from the interrupt handler, the
instruction counter is incremented erroneously by one. As this
behaviour is deterministic, it can be compensated for by subtracting
the number of switches to/from supervisor mode (this event can also be
counted in hardware).

Interesting features of the MPC are to configure the counters to only
increment in user and/or supervisor mode and only count instructions
executed by a specially marked process. This way it is possible to
only count instructions executed by the UML virtual machine. When the
VM accesses services of the host OS, these requests are performed in
kernel mode (after a system call) and the instructions are not
counted. This is essential because the host operating system gives no
guarantees that it will execute the same number of instructions during
the replay. Similarly any paging operation performed by the host OS
are performed in kernel mode and thus transparent to the instruction
counter of the UML.

During the replay the so-called performance monitor interrupt is
used. It creates (physical) interrupts after a pre-recorded number of
executed instructions. By diverting these interrupts to special replay
handlers in the UML, its interrupts and signals can be replayed.

Besides recording the moment of interrupts, also input from external
devices needs to be stored and replayed. For UML this done is in a
straightforward manner by extending the (virtual) device drivers. The
attractivity of using a VM for execution replay is mostly due to the
fact, that this driver interface is small and well-defined.

The described setup allows to record and reproduce the virtual
machine's behaviour, including its applications. In order to debug an
application a modified debugger is needed, that attaches to a process
{\it within} the UML VM. Normally a debugger is supported by the
operating system to attach to a process. In this case the {\it guest}
OS cannot possibly be aware of the debugger, as it merely replays a
recorded instruction sequence.  Thus the debugger (executing in the
host OS) must be aware of the structure and memory layout of the guest
OS in order to read the stacks, register contents, set breakpoints and
single step through the code etc. It is intended to adapt {\tt gdb} or
a kernel debugger ({\tt kgdb}) for this purpose.

\section{Outlook}
\label{sec:outlook}

To be generically applicable, the overhead in the execution phase must
be small enough to execute the program with sufficient speed. Many
application programs --- in particular GUI driven software --- have
very moderate performance requirements and run satisfactorily on
low-end machines. Therefore the author assumes that a high-end machine
will have enough (processor) resources to record the input stimuli in
the background without slowing down the program too much. If this
assumption holds, the chosen approach will allow to apply the ER
concept to nearly all areas of software development --- without
modifications to the program or system.

As this paper illustrates, the base technology for making ER a reality
is available today: a generic virtual machine environment and software
or hardware instruction counters. It is the aim of the author to
combine these technologies for further research in this area and to
prove its viability.

In the long run, improved hardware instruction counters may allow
simplified implementations in more architectures. Also special
hardware extensions dedicated solely to execution replay are
conceivable. These may include support for virtual machines (as known
from mainframe systems like the IBM S390) or automatically record data
read from peripheral registers or save time stamps for interrupts. By
careful design, such hardware extensions might even perform all
recording stealthily in the background and thus eliminate the
so-called probe effect. This would allow to apply ER to all software
executing on such a machine.

\label{end}

\bibliography{final}

\begin{thebibliography}{RBdK00}

\bibitem[BG91]{bacon91hardwareassisted}
David~F. Bacon and Seth~Copen Goldstein.
\newblock Hardware-assisted replay of multiprocessor programs.
\newblock {\em Proceedings of the ACM/ONR Workshop on Parallel and Distributed
  Debugging, published in ACM SIGPLAN Notices}, 26(12):194--206, 1991.

\bibitem[Dik]{www-dike}
Jeff Dike.
\newblock {User Mode Linux}.
\newblock http://user-mode-linux.sourceforge.net/.

\bibitem[Hus02]{huselius_2002}
Joel Huselius.
\newblock {D}ebugging {P}arallel {S}ystems: {A} {S}tate of the {A}rt {R}eport.
\newblock Technical Report~63, M\"{a}lardalen University, Department of
  Computer Science and Engineering, September 2002.

\bibitem[RBdK00]{ronsse00execution}
Michiel Ronsse, Koenraad~De Bosschere, and Jacques~Chassin de~Kergommeaux.
\newblock Execution replay and debugging.
\newblock In {\em Automated and Algorithmic Debugging}, 2000.

\bibitem[TH00]{thane00}
Henrik Thane and Hans Hansson.
\newblock {Using Deterministic Replay for Debugging of Distributed RealTime
  Systems}.
\newblock In the 12th Euromicro Conference on Real- Time Systems, pages 265 -
  272. IEEE Computer Society, June 2000.

\bibitem[TPS]{thane01asterix}
H.~Thane, A.~Pettersson, and D.~Sundmark.
\newblock {The Asterix Realtime Kernel. In proceedings of the 13th Euromicro
  Conference on Real-Time Systems (ECRTS'01), Industrial Session, Delft, June
  2001}.
\newblock http://citeseer.nj.nec.com/thane01asterix.html.

\bibitem[{VMw}]{www-vmware}
{VMware Inc.}
\newblock http://www.vmware.com/.

\end{thebibliography}

\end{document}